# Unexplainability and Incomprehensibility of Artificial Intelligence

**Roman V. Yampolskiy**
Computer Engineering and Computer Science
University of Louisville
roman.yampolskiy@louisville.edu, @romanyam
June 20, 2019

> "*If a lion could speak, we couldn't understand him*"
> Ludwig Wittgenstein

> "*It would be possible to describe everything scientifically, but it would make no sense. It would be a description without meaning - as if you described a Beethoven symphony as a variation of wave pressure.*"
> Albert Einstein

> "*Some things in life are too complicated to explain in any language. ... Not just to explain to others but to explain to yourself. Force yourself to try to explain it and you create lies.*"
> Haruki Murakami

> "*I understand that you don't understand*"
> Grigori Perelman

**Abstract**
Explainability and comprehensibility of AI are important requirements for intelligent systems deployed in real-world domains. Users want and frequently need to understand how decisions impacting them are made. Similarly it is important to understand how an intelligent system functions for safety and security reasons. In this paper, we describe two complementary impossibility results (Unexplainability and Incomprehensibility), essentially showing that advanced AIs would not be able to accurately explain some of their decisions and for the decisions they could explain people would not understand some of those explanations.

**Keywords:** *AI Safety, Black Box, Comprehensible, Explainable AI, Impossibility, Intelligible, Interpretability, Transparency, Understandable, Unserveyability.*

## 1. Introduction

For decades AI projects relied on human expertise, distilled by knowledge engineers, and were both explicitly designed and easily understood by people. For example, expert systems, frequently based on decision trees, are perfect models of human decision making and so are naturally understandable by both developers and end-users. With paradigm shift in the leading AI methodology, over the last decade, to machine learning systems based on Deep Neural Networks (DNN) this natural ease of understanding got sacrificed. The current systems are seen as "black boxes" (not to be confused with AI boxing [1, 2]), opaque to human understanding but extremely capable both with respect to results and learning of new domains. As long as Big Data and Huge Compute are available, zero human knowledge is required [3] to achieve superhuman [4] performance.

With their new found capabilities DNN-based AI systems are tasked with making decisions in employment [5], admissions [6], investing [7], matching [8], diversity [9], security [10, 11], recommendations [12], banking [13], and countless other critical domains. As many such domains are legally regulated, it is a desirable property and frequently a requirement [14, 15] that such systems should be able to explain how they arrived at their decisions, particularly to show that they are bias free [16]. Additionally, and perhaps even more importantly to make artificially intelligent systems safe and secure [17] it is essential that we understand what they are doing and why. A particular area of interest in AI Safety [18-25] is predicting and explaining causes of AI failures [26].

A significant amount of research [27-41] is now being devoted to developing explainable AI. In the next section we review some main results and general trends relevant to this paper.

## 2. Literature Review
Hundreds of papers have been published on eXplainable Artificial Intelligence (XAI) [42]. According to DARPA [27], XAI is supposed to "produce more explainable models, while maintaining a high level of learning performance … and enable human users to understand, appropriately, trust, and effectively manage the emerging generation of artificially intelligent partners". Detailed analysis of literature on explainability or comprehensibility is beyond the scope of this paper, but the readers are encouraged to look at many excellent surveys of the topic [43-45]. Miller [46] surveys social sciences to understand how people explain, in the hopes of transferring that knowledge to XAI, but of course people often say: "I can't explain it" or "I don't understand". For example, most people are unable to explain how they recognize faces, a problem we frequently ask computers to solve [47, 48].

Despite wealth of publications on XAI and related concepts [49-51], the subject of unexplainability or incomprehensibility of AI is only implicitly addressed. Some limitations of explainability are discussed: "ML algorithms intrinsically consider high-degree interactions between input features, which make disaggregating such functions into human understandable form difficult. … While a single linear transformation may be interpreted by looking at the weights from the input features to each of the output classes, multiple layers with non-linear interactions at every layer imply disentangling a super complicated nested structure which is a difficult task and potentially even a questionable one [52]. … As mentioned before, given the complicated structure of ML models, for the same set of input variables and prediction targets, complex machine learning algorithms can produce multiple accurate models by taking very similar but not the same internal pathway in the network, so details of explanations can also change across multiple accurate models. This systematic instability makes automated generated explanations difficult." [42].

Sutcliffe et al. talk about incomprehensible theorems [53]: "Comprehensibility estimates the effort required for a user to understand the theorem. Theorems with many or deeply nested structures may be considered incomprehensible." Muggleton et al. [54] suggest "using inspection time as a proxy for incomprehension. That is, we might expect that humans take a long time … in the case they find the program hard to understand. As a proxy, inspection time is easier to measure than comprehension."

The tradeoff between explainability and comprehensibility is recognized [52], but is not taken to its logical conclusion. "[A]ccuracy generally requires more complex prediction methods [but] simple and interpretable functions do not make the most accurate predictors" [55]. "Indeed, there are algorithms that are more interpretable than others are, and there is often a tradeoff between accuracy and interpretability: the most accurate AI/ML models usually are not very explainable (for example, deep neural nets, boosted trees, random forests, and support vector machines), and the most interpretable models usually are less accurate (for example, linear or logistic regression)." [42].

Incomprehensibility is supported by well-known impossibility results. Charlesworth proved his Comprehensibility theorem while attempting to formalize the answer to such questions as: "If [full human-level intelligence] software can exist, could humans understand it?" [56]. While describing implications of his theorem on AI, he writes [57]: "Comprehensibility Theorem is the first mathematical theorem implying the impossibility of any AI agent or natural agent—including a not-necessarily infallible human agent—satisfying a rigorous and deductive interpretation of the self-comprehensibility challenge. … Self-comprehensibility in some form might be essential for a kind of self-reflection useful for self-improvement that might enable some agents to increase their success." It is reasonable to conclude that a system which doesn't comprehend itself would not be able to explain itself.

Hernandez-Orallo et al. introduce the notion of K-incomprehensibility (a.k.a. K-hardness) [58]. "This will be the formal counterpart to our notion of hard-to-learn good explanations. In our sense, a k-*incomprehensible* string with a high *k* (difficult to comprehend) is different (harder) than a k-*compressible* string (difficult to learn) [59] and different from classical computational complexity (slow to compute). Calculating the value of *k* for a given string is not computable in general. Fortunately, the converse, i.e., given an arbitrary *k*, calculating whether a string is k-*comprehensible* is computable. … Kolmogorov Complexity measures the amount of information but not the complexity to understand them." [58].

Yampolskiy addresses limits of understanding other agents in his work on the space of possible minds [60]: "Each mind design corresponds to an integer and so is finite, but since the number of minds is infinite some have a much greater number of states compared to others. This property holds for all minds. Consequently, since a human mind has only a finite number of possible states, there are minds which can never be fully understood by a human mind as such mind designs have a much greater number of states, making their understanding impossible as can be demonstrated by the pigeonhole principle." Hibbard points out safety impact from incomprehensibility of AI: "Given the incomprehensibility of their thoughts, we will not be able to sort out the effect of any conflicts they have between their own interests and ours."

We are slowly starting to realize that as AIs become more powerful, the models behind their success will become ever less comprehensible to us [61]: "… *deep learning* that produces outcomes based on so many different variables under so many different conditions being transformed by so many layers of neural networks that humans simply cannot comprehend the model the computer has built for itself. … Clearly our computers have surpassed us in their power to discriminate, find patterns, and draw conclusions. That's one reason we use them. Rather than reducing phenomena to fit a relatively simple model, we can now let our computers make models as big as they need to. But this also seems to mean that what we know depends upon the output of machines the functioning of which we cannot follow, explain, or understand. … But some of the new models are incomprehensible. They can exist only in the weights of countless digital triggers networked together and feeding successive layers of networked, weighted triggers representing huge quantities of variables that affect one another in ways so particular that we cannot derive general principles from them."

"Now our machines are letting us see that even if the rules are simple, elegant, beautiful and rational, the domain they govern is so granular, so intricate, so interrelated, with everything causing everything else all at once and forever, that our brains and our knowledge cannot begin to comprehend it. … Our new reliance on inscrutable models as the source of the justification of our beliefs puts us in an odd position. If knowledge includes the justification of our beliefs, then knowledge cannot be a class of mental content, because the justification now consists of models that exist in machines, models that human mentality cannot comprehend. … But the promise of machine learning is that there are times when the machine's inscrutable models will be far more predictive than the manually constructed, human-intelligible ones. In those cases, our knowledge—if we choose to use it—will depend on justifications that we simply cannot understand. … [W]e are likely to continue to rely ever more heavily on justifications that we simply cannot fathom. And the issue is not simply that we cannot fathom them, the way a lay person can't fathom a string theorist's ideas. Rather, it's that the nature of computer-based justification is not at all like human justification. It is alien." [61].

## 3. Unexplainability

A number of impossibility results are well-known in many areas of research [62-70] and some are starting to be discovered in the domain of AI research, for example: Unverifiability [71], Unpredictability[1] [72] and limits on preference deduction [73] or alignment [74]. In this section we introduce *Unexplainability* of AI and show that some decisions of superintelligent systems will never be explainable, even in principle. We will concentrate on the most interesting case, a superintelligent AI acting in novel and unrestricted domains. Simple cases of Narrow AIs making decisions in restricted domains (Ex. Tic-Tac-Toe) are both explainable and comprehensible. Consequently a whole spectrum of AIs can be developed from completely explainable/comprehensible to completely unexplainable/incomprehensible. We define Unexplainability as impossibility of providing an explanation for certain decisions made by an intelligent system which is both 100% accurate and comprehensible.

Artificial Deep Neural Networks continue increasing in size and may already comprise millions of neurons, thousands of layers and billions of connecting weights, ultimately targeting and perhaps surpassing the size of the human brain. They are trained on Big Data from which million feature vectors are extracted and on which decisions are based, with each feature contributing to the decision in proportion to a set of weights. To explain such a decision, which relies on literally billions of contributing factors, AI has to either simplify the explanation and so make the explanation less accurate/specific/detailed or to report it exactly but such an explanation elucidates nothing by virtue of its semantic complexity, large size and abstract data representation. Such precise reporting is just a copy of trained DNN model.

---

[1] Unpredictability is not the same as Unexplainability or Incomprehensibility, see ref. 72. Yampolskiy, R.V., *Unpredictability of AI.* arXiv preprint arXiv:1905.13053, 2019. for details.

For example, an AI utilized in the mortgage industry may look at an application to decide credit worthiness of a person in order to approve them for a loan. For simplicity, let's say the system looks at only a hundred descriptors of the applicant and utilizes a neural network to arrive at a binary approval decision. An explanation which included all hundred features and weights of the neural network would not be very useful, so the system may instead select one of two most important features and explain its decision with respect to just those top properties, ignoring the rest. This highly simplified explanation would not be accurate as the other 98 features all contributed to the decision and if only one or two top features were considered the decision could have been different. This is similar to how Principal Component Analysis works for dimensionality reduction [75].

Even if the agent trying to get the explanation is not a human but another AI the problem remains as the explanation is either inaccurate or agent-encoding specific. Trained model could be copied to another neural network, but it would likewise have a hard time explaining its decisions. Superintelligent systems not based on DNN would face similar problems as their decision complexity would be comparable to those based on neural networks and would not permit production of efficient and accurate explanations. The problem persists in the case of self-referential analysis, where a system may not understand how it is making a particular decision.

Any decision made by the AI is a function of some input data and is completely derived from the code/model of the AI, but to make it useful an explanation has to be simpler than just presentation of the complete model while retaining all relevant, to the decision, information. We can reduce this problem of explaining to the problem of lossless compression [76]. Any possible decision derived from data/model can be represented by an integer encoding such data/model combination and it is a proven fact that some random integers can't be compressed without loss of information due to the Counting argument [77]. "The pigeonhole principle prohibits a bijection between the collection of sequences of length *N* and any subset of the collection of sequences of length *N* - 1. Therefore, it is not possible to produce a lossless algorithm that reduces the size of every possible input sequence."[2] To avoid this problem, an AI could try to produce decisions, which it knows are explainable/compressible, but that means that it is not making the best decision with regards to the given problem, doing so is suboptimal and may have safety consequences and so should be discouraged.

Overall, we should not be surprised by the challenges faced by Artificial Neural Networks attempting to explain their decision, as they are modeled on Natural Neural Networks of human beings and people are also "black boxes" as illustrated by a number of split brain experiments [78]. In such experiments it is frequently demonstrated that people simply make up explanations for their actions after the decision has already been made. Even to ourselves, we rationalize our decision after the fact and don't become aware of our decisions or how we made them until after they been made unconsciously [79]. People are notoriously bad at explaining certain decisions such as how they recognize faces or what makes them attracted to a particular person. These reported limitations in biological agents support idea that unexplainability is a universal impossibility result impacting all sufficiently complex intelligences.

---

[2] https://en.wikipedia.org/wiki/Lossless_compression

## 4. Incomprehensibility

A complimentary concept to Unexplainability, *Incomprehensibility* of AI address capacity of people to completely understand an explanation provided by an AI or superintelligence. We define Incomprehensibility as an impossibility of completely understanding any 100% - accurate explanation for certain decisions of intelligent system, by any human.

Artificially intelligent systems are designed to make good decision in their domains of deployment. Optimality of the decision with respect to available information and computational resources is what we expect from a successful and highly intelligent systems. An explanation of the decision, in its ideal form, is a proof of correctness of the decision. (For example, a superintelligent chess playing system may explain why it sacrificed a queen by showing that it forces a checkmate in 12 moves, and by doing so proving correctness of its decision.) As decisions and their proofs can be arbitrarily complex impossibility results native to mathematical proofs become applicable to explanations. For example, explanations may be too long to be surveyed [80, 81] (Unserveyability), Unverifiable [71] or too complex to be understood [82] making the explanation incomprehensible to the user. Any AI, including black box neural networks can in principle be converted to a large decision tree of nothing but "if" statements, but it will only make it human-readable not human-understandable.

It is generally accepted that in order to understand certain information a person has to have a particular level of cognitive ability. This is the reason students are required to take standardized exams such as SAT, ACT, GRE, MCAT or LCAT, etc. and score at a particular percentile in order to be admitted to their desired program of study at a selective university. Other, but similar tests are given to those wishing to join the military or government service. All such exams indirectly measure person's IQ (Intelligence Quotient) [83, 84] but vary significantly in how closely they correlate with standard IQ test scores (g-factor loading). The more demanding the program of study (even at the same university), the higher cognitive ability is expected from students. For example, average quantitative GRE score of students targeting mathematical sciences is 163, while average quantitative score for students interested in studying history is 148[3]. The trend may be reversed for verbal scores.

People often find themselves in situations where they have to explain concepts across a significant communication range [85] for example to children or to people with mental challenges. The only available option in such cases is to provide a greatly oversimplified version of the explanation or a completely irrelevant but simple explanation (a lie). In fact the situation is so common we even have a toolbox of common "explanations" for particular situations. For example, if a five-year old asks: "Where do babies come from?" They are likely to hear something like "A seed from the daddy and an egg from the mommy join together in the mom's tummy"[4], instead of a talk about DNA, fertilization and womb. A younger child may learn that the "stork brings them" or "they come from a baby store". Alternatively, an overly technical answer could be provided to confuse the child into thinking they got an answer, but with zero chance of them understanding such overcomplicated response. Overall, usefulness of an explanation is relative to the person who is trying to comprehend it. The same explanation may be comprehended by one person, and completely misunderstood by another.

---

[3] https://www.prepscholar.com/gre/blog/average-gre-scores/

There is a similar and perhaps larger intelligence gap between superintelligence and adult humans, making the communication range unsurmountable. It is likely easier for a scientist to explain quantum physics to a mentally challenged deaf and mute four-year-old raised by wolves then for superintelligence to explain some of its decisions to the smartest human. We are simply not smart enough to understand certain concepts. Yampolskiy proposed [82] a complexity measure which is based on the minimum intelligence necessary to understand or develop a particular algorithm, and while it takes less intelligence to just understand rather than create both requirements could be well above IQ of the smartest human. In fact it could be very hard to explain advanced concepts to even slightly less intelligent agents.

We can predict a certain complexity barrier to human understanding for any concept for which relative IQ of above 250 would be necessary, as no person has ever tested so high. In practice the barrier may be much lower, as average IQ is just 100 and additional complication from limited memory and limited attention spans can place even relative easy concepts outside of human grasp. To paraphrase Wittgenstein: if superintelligence explained itself we would not understand it.

Given that research on deception by AI is well established [86] it would not be difficult for advanced AIs to provide highly believable lies to their human users. In fact such explanations can be designed to take advantage of AI's knowledge of the human behavior [87, 88] and mental model [89, 90], and manipulate users beyond just convincing them that explanation is legitimate [91]. AI would be able to target explanations to the mental capacity of particular people, perhaps taking advantage of their inherent limitations. It would be a significant safety issue, and it is surprising to see some proposals for using human users as targets of competing (adversarial) explanations from AIs [92].

Incomprehensibility results are well-known for different members of Chomsky hierarchy [93] with finite state automation unable to recognize context-free languages, pushdown automata unable to recognize context-sensitive languages and linear-bounded non-deterministic Turing machines unable to recognize recursively enumerable languages. Simpler machines can't recognize languages which more complex machines can recognize.

While people are frequently equated with unrestricted Turing machines via Church-Turing thesis [94], Blum et al. formalize human computation, in practice, as a much more restricted class [95]. However, Turing machines are not an upper limit on what is theoretically computable as described by different hypercomputation models [96]. Even if our advanced AIs (superintelligence), fail to achieve true hypercomputation capacity, for all practical purposes and compared to the human computational capabilities they would be outside of what human-equivalent agents can recognize/comprehend.

---

[4] https://www.babycenter.com/0_how-to-talk-to-your-child-about-sex-age-5_67112.bc

Superintelligence would be a different type of computation, far superior to humans in practice. It is obviously not the case that superintelligent machines would actually have infinite memories or speeds but they would appear to act as they do to unaugmented humans. For example a machine capable of remembering one trillion items vs seven, in short-term memory of people, would appear to have infinite capacity to memorize. In algorithmic complexity theory some algorithms become the most efficient for a particular problem type on inputs so large as to be unusable in practice, but such inputs are nonetheless finite [97]. So, just like a finite state automata can't recognize recursively enumerable languages, so will people fail in practice to comprehend some explanations produced by superintelligent systems, they are simply not in the same class of automata, even if theoretically, given infinite time, they are.

Additionally, decisions made by AI could be mapped onto the space of mathematical conjectures about the natural numbers. An explanation for why a particular mathematical conjecture is true or false would be equivalent to a proof (for that conjecture). However, due to Gödel's First Incompleteness Theorem we know that some true conjectures are unprovable. As we have mapped decision on conjectures and explanations on proofs, that means that some decision made by AI are fundamentally unexplainable/incomprehensible. Explanations as proofs would be subject to all the other limitations known about proofs, including Unserveyability, Unverifiability and Undefinability [98, 99]. Finally, it is important to note that we are not saying that such decision/conjecture reduction would preserve semantics of the subject, just that it is a useful tool for showing impossibility of explainability/comprehensibility in some cases.

## 5. Conclusions

The issues described in this paper can be seen as a communication problem between AI encoding and sending information (sender) and person receiving and decoding information (receiver). Efficient encoding and decoding of complex symbolic information is difficult enough, as described by Shannon's Information Theory [100], but with Explainability and Comprehensibility of AI we also have to worry about complexity of semantic communication [101]. Explainability and Comprehensibility are another conjugate pair [71, 102] in the domain of AI safety. The more accurate is the explanation the less comprehensible it is, and vice versa, the more comprehensible the explanation the less accurate it is. A non-trivial explanation can't be both accurate and understandable, but it can be inaccurate and comprehensible. There is a huge difference between understanding something and almost understanding it. Incomprehensibility is a general result applicable to many domains including science, social interactions, etc. depending on a mental capacity of a participating person(s).

Human being are finite in our abilities. For example our short term memory is about 7 units on average. In contrast, an AI can remember billions of items and their capacity to do so grows exponentially, while never infinite in a true mathematical sense, machine capabilities can be considered such in comparison to ours. This is true for memory, compute speed and communication abilities. Hence the famous: *Finitum Non Capax Infiniti* (The finite cannot contain the infinite) is highly applicable to understand the incomprehensibility of the god-like [103] superintelligent AIs.

Shown impossibility results present a number of problems for AI Safety. Evaluation and debugging of intelligent systems becomes much harder if their decisions are unexplainable/incomprehensible. In particular, in case of AI failures [104] accurate explanations are necessary to understand the problem and reduce likelihood of future accidents. If all we have is a "black box" it is impossible to understand causes of failure and improve system safety. Additionally, if we grow accustomed to accepting AI's answers without an explanation, essentially treating it as an Oracle system, we would not be able to tell if it begins providing wrong or manipulative answers.

## Acknowledgments

The author is grateful to Elon Musk and the Future of Life Institute, and to Jaan Tallinn and Effective Altruism Ventures for partially funding his work on AI Safety. It is obvious that the subject described in this paper is accurately explained and completely comprehensible.